# Bigger Is Not Better: The Fastest Static GPU Index Is Also Lightweight!

Justus Henneberg
henneberg@uni-mainz.de
Johannes Gutenberg University
Mainz, Germany

Felix Schuhknecht
schuhknecht@uni-mainz.de
Johannes Gutenberg University
Mainz, Germany

## Abstract

Sorting and binary searching a dense array can be considered the simplest and most space efficient form of indexing. This holds especially on GPUs as they exhibit exceptional sorting performance. However, the popular opinion is that such a primitive approach cannot compete with large, highly-sophisticated GPU index structures in terms of lookup performance, and hence, should not actually be considered in practice.

In this work, we will investigate whether binary search actually still deserves this bad reputation or whether it can be a fast and space-minimal alternative to more heavy-weight index structures, in particular when utilizing all the advancements of current highly-parallel GPU architectures. To find out, we introduce advanced variants of binary search to GPUs and equip them with a set of established low-level optimizations. These architecture-specific optimizations aim at getting the most out of binary search by (a) greatly reducing the overall amount of GPU memory accesses required during search, (b) exploiting the enormous benefits of memory access coalescing on a GPU, and (c) maximizing scalability by reordering the dataset into a more favorable layout.

By comparing our optimized search strategies against nine state-of-the-art GPU index structures under several static indexing workloads, we demonstrate that they not only outperform all competitors (except for hashing-based approaches) by a factor of up to 3.8, but also maintain the smallest possible memory footprint.

## 1 Introduction

Database indexing generally comes down to two main tasks: Given a table column $C$, (1) find the row ID of the row that contains a fixed value $X$ in $C$ (*point lookup*) and (2) find the row IDs of all rows where the entry $e$ in $C$ satisfies $L \leq e \leq U$ for fixed values $L, U$ (*range lookup*). While there exist various highly sophisticated index structures that solve this problem in one way or the other, the most straight-forward way is obviously copying each entry of $C$ along with its unique row ID into a separate buffer, re-arranging the entries in ascending order, and then performing binary search to answer point or range lookups. The beauty of this approach lies not only its simplicity, but more importantly in its minimal memory footprint, which is a very important property in the presence of scarce GPU memory.

Unfortunately, doubts about its performance in comparison to more heavy-weight index structures have always hindered the adoption of binary search in practice, even in read-only or read-mostly scenarios such as index-nested-loop joins. However, we ask whether the doubts about binary search are in fact still reasonable — and whether they even ever were — based on the following two observations: First, there is no single text-book binary search, but there exist numerous different variants of the concept [12, 24, 31, 32], each applying various different optimizations such as prefetching, exploiting SIMD instructions, or reordering the dataset somehow to achieve a better cache efficiency. Second, most findings about binary search originate from the CPU world and/or are rather outdated by now due to the old age of the method. The consequence of both is that it remains unclear how the individual variants of the method actually perform on modern highly-parallel GPU architectures. Recent work on GPU-resident index structures [18, 19] showed that binary search performs surprisingly well on certain workloads, indicating that the situation might actually be different than it used to be.

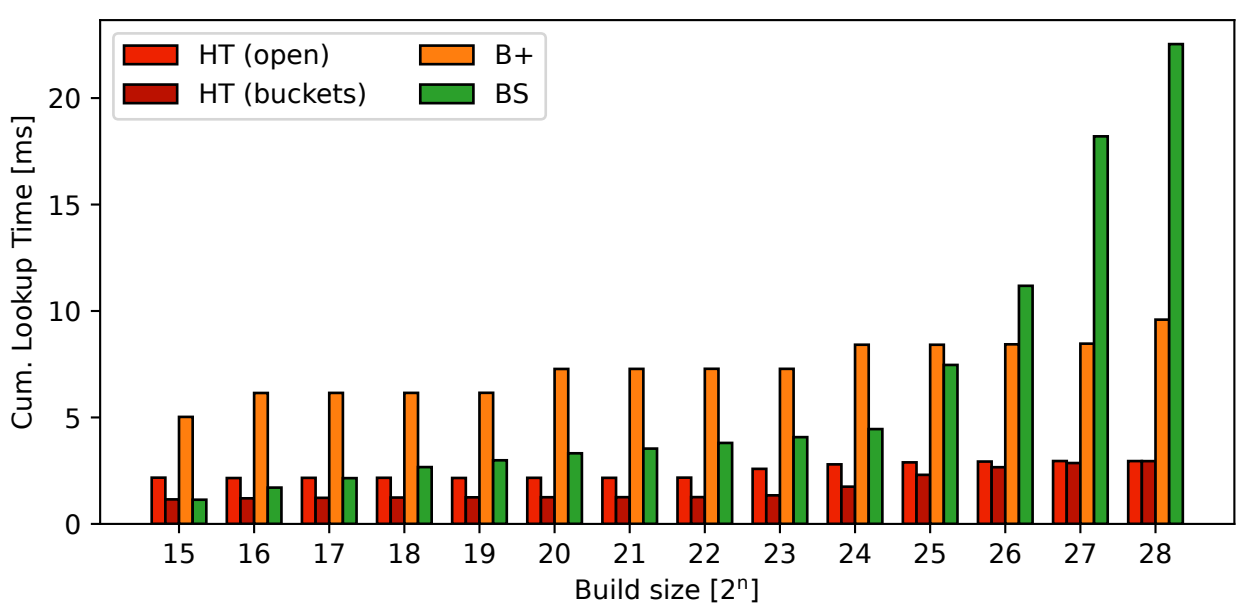

**Figure 1: Binary search outperforms traditional index structures on small build sets.**

To get a first impression of binary search on GPUs, let us inspect in Figure 1 the cumulative lookup time of the text-book binary search (**BS**) in comparison with three state-of-the-art GPU indexes, namely an open-addressing hash table (**HT (open)**) [21, 23], a bucket-based hash table (**HT (buckets)**) [4] and a B+-Tree (**B+**) [7, 9] while varying the build set size. While the performance of all indexes increases when the dataset becomes smaller, **BS** interestingly benefits far more than the remaining indexes, even starting to outperform **B+** and **HT** for very small sizes. We attribute this to the low code complexity of binary search, which results in fewer instructions being executed. Further, with its minimal memory footprint, it can remain fully cache-resident longer than its competitors, which consume more space due to building auxiliary data structures (**B+**) or having to over-allocate memory to yield reasonable performance (**HT**). However, we can also see that for larger build sizes, the text-book **BS** is unfortunately not competitive. To find out whether the picture is different for advanced variants of binary search, we make the following contributions in this work:

**(1)** We first introduce a variant of binary search called *Eytzinger binary search* (**EBS**) to GPUs, as it has proven itself on CPUs. Eytzinger does not order the entries by key, but orders them with respect to the access pattern of the search paths. Since **EBS** requires a special element order, we also present a novel parallel algorithm to construct this order using only one read and one write operation per thread, which is the theoretical optimum. Further, we propose a method to perform range lookups efficiently despite the element order for **EBS** not being strictly ascending.

**(2)** We then generalize **EBS** to a $k$-ary variant called *Eytzinger $k$-ary search* (**EKS**), which maintains the minimal footprint of **EBS** as well as its feature set, but offers more flexibility regarding the search fan-out and hence the parallelization capabilities. Consequently, **EKS** can adapt to workload changes and overall better utilize the available memory bandwidth and processing power.

**(3)** On top of both **EBS** and **EKS**, we integrate the support for a set of four promising optimizations, namely *limiting concurrency*, *cache pinning*, *lookup reordering*, and different *memory layouts* to be able to study their positive or negative impact on both variants. While these optimization have partially been studied before, it remains unclear how they perform on current GPUs and in combination with our specific variants.

**(4)** Before comparing **EBS** and **EKS** against state-of-the-art indexes, we tune both methods to their limits and identify all inherent parameter choice trade-offs, since both variants as well as the optimizations are highly configurable. As a result, we identify four strong configurations that we use in the following evaluation.

**(5)** To identify whether and in which situations **EBS** and **EKS** actually outperform sophisticated indexes, we perform a rich set of experiments against nine baselines, including multiple hash tables, a learned index, and a ray-tracing based index. We compare the capabilities of all methods in terms of point and range lookups, build time, memory footprint, behavior under skew, pre-sorted lookups, 64-bit keys, duplicate keys, and different GPU architectures. As a core result, we identify that our binary search variants outperform all competitors (except for hashing-based approaches) by a factor of up to 3.8 *while* maintaining the smallest possible memory footprint.

## 2 Hardware Architecture

Before diving into index design, let us discuss the architecture and specifics of the underlying hardware. As our main testbed, we use an NVIDIA RTX 5090 (Blackwell architecture), a high-end consumer GPU. The architecture is visualized in Figure 2. Processing on the 5090 is distributed across 170 *streaming multiprocessors* (SMs) containing 4 vector units of 32 cores each. Typically, one unit of work is assigned to one thread. When running a multi-threaded task, the GPU distributes small batches of threads (called *thread blocks*) across SMs, which then schedule even smaller batches of 32 threads (called *warps*) onto the 32-core vector units. Each SM has its own shared L1 cache (128 KB), which can be partially (up to 100 KB) reconfigured to serve as fast user-programmable scratch memory, allowing for fast data movement and storage within a thread block. Similarly, each SM contains 256 KB worth of 32-bit registers, which can be cross-referenced by threads within a warp to exchange data on-the-fly (called *warp shuffling*). Additionally, the GPU has a global L2 cache (100 MB). To ensure scalability, NVIDIA restricts direct communication across SMs, only allowing indirect communication though the GPU's globally-accessible main memory. Similar to block storage devices, loading large contiguous chunks from main memory is more efficient than small random accesses. Consequently, memory throughput is maximized if neighboring threads in a warp access neighboring memory addresses, typically referred to as a *coalesced memory access.* We will show experimental evidence for this in Section 5.

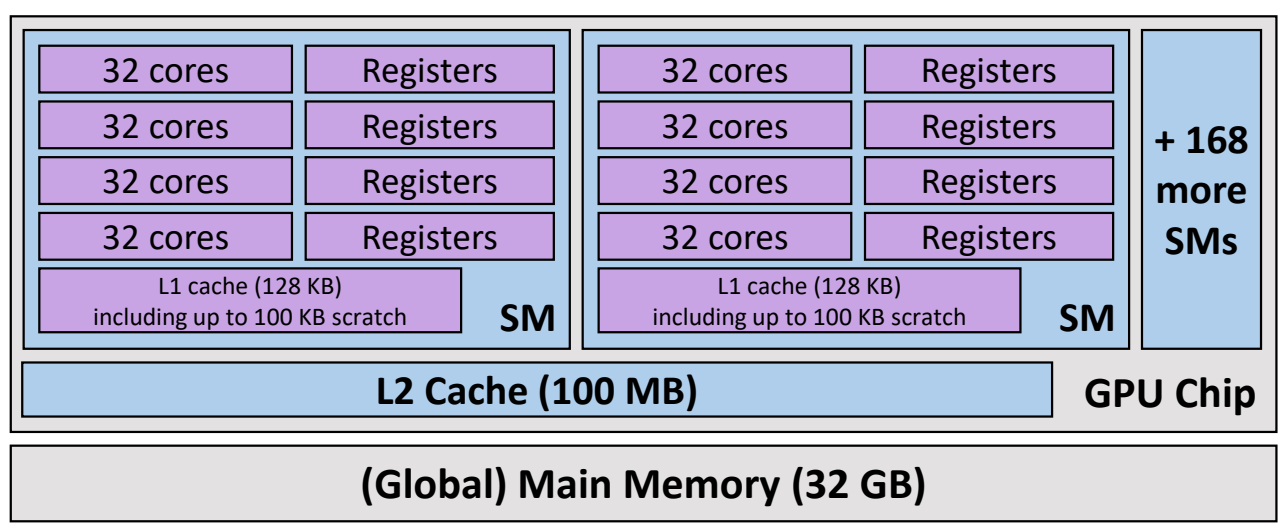

**Figure 2: Architecture of the NVIDIA RTX 5090.**

## 3 From Binary Search To Eytzinger Search

We briefly recap how a traditional left-or-right binary search finds a target key $X$ in a sorted array. We assume $n = 2^m - 1$ for some integer $m$, for reasons that will become obvious later on, and generalize to arbitrary $n$ afterwards. If the entry in the center (called *pivot*) of the array matches $X$, the search terminates. Otherwise, we apply this procedure iteratively in either the left or the right half, both of which are of size $2^{k-1} - 1$. This algorithm performs only a logarithmic number of memory accesses to find $X$, and building the structure just requires a key-value sort, which is a highly-optimized operation on GPUs and already a part of many GPU libraries, such as NVIDIA's CUB [14] or Thrust[15].

Figure 3 visualizes the possible search paths for an example dataset with $n = 15 = 2^4 - 1$ entries. The number below each entry corresponds to the number of search steps after which the element is first visited. The search paths can also be visualized as a balanced binary tree, such as the one in Figure 4.

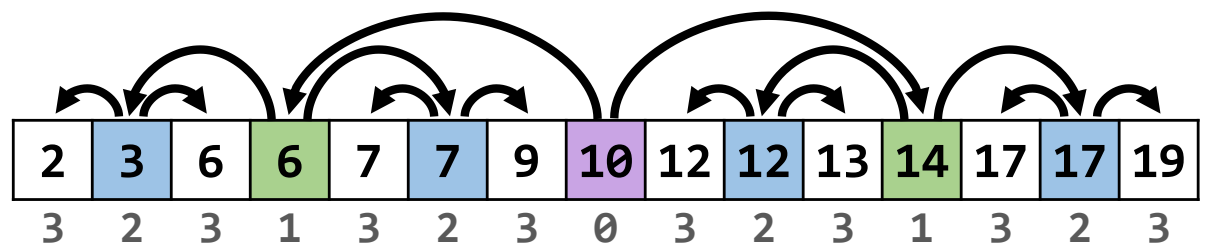

**Figure 3: Search paths of a traditional left-or-right binary search. The numbers denote the search tree level of each entry.**

Looking at all possible paths, it becomes obvious that only a small subset of the entries is accessed during the first few steps. However, storing the array in ascending order places these elements far apart in memory, resulting in poor cache efficiency during traversal. A solution to this problem is to rearrange the keys so that every level of the tree is stored as one contiguous chunk, and the levels are stored one after the other, following the order of traversal, called

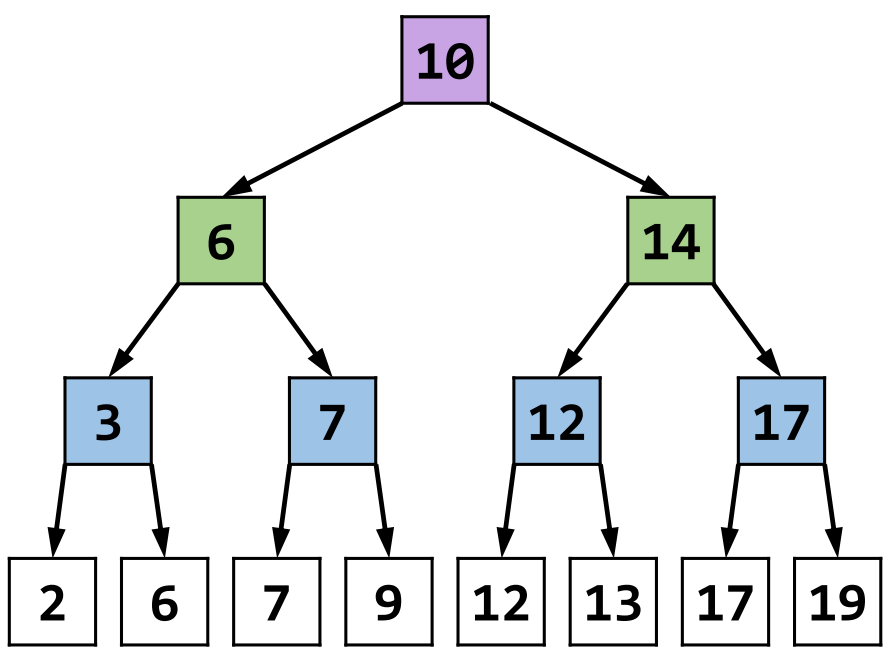

**Figure 4: Visualization of the search path of left-or-right binary search as a balanced binary tree.**

*linearization.* This way, the entries accessed during the first few steps are distributed across fewer cache lines, which results in (1) fewer cache misses across subsequent steps and (2) fewer cache misses if two threads traverse the structure concurrently. Whenever $n = 2^m - 1$, we can easily represent the binary tree from Figure 4 as a dense array with implicit parent-child relations, similar to a binary heap. The resulting dense representation is generally referred to as *Eytzinger order* [25] and shown in Figure 5. Note that there is an extra empty slot at the beginning, as this results in even better cache alignment. We will present the exact build algorithm for arbitrary $n$ in Section 4.

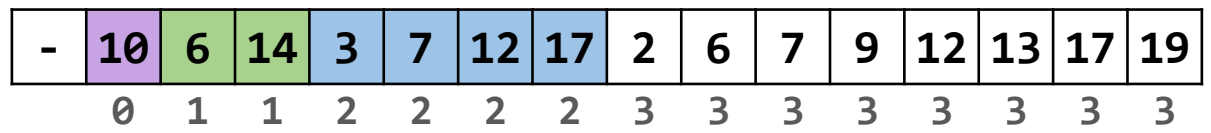

**Figure 5: Eytzinger order for our example dataset.**

An Eytzinger binary search (**EBS**) works similar to the traditional left-or-right binary search: We start our search at position 1 (because slot 0 is always empty). Given an entry at position $i$, we proceed to position $2i$ if the entry is greater than or equal to the target key, and to position $2i + 1$ if it is smaller. The search fails once $i$ exceeds the array's size. We visualize the difference between the search procedures in Figure 6. The top part of the figure shows the search path of the traditional left-or-right binary search on our example key set. The bottom part shows the same search with keys rearranged into Eytzinger order. The arrows indicate the steps taken to locate key 9.

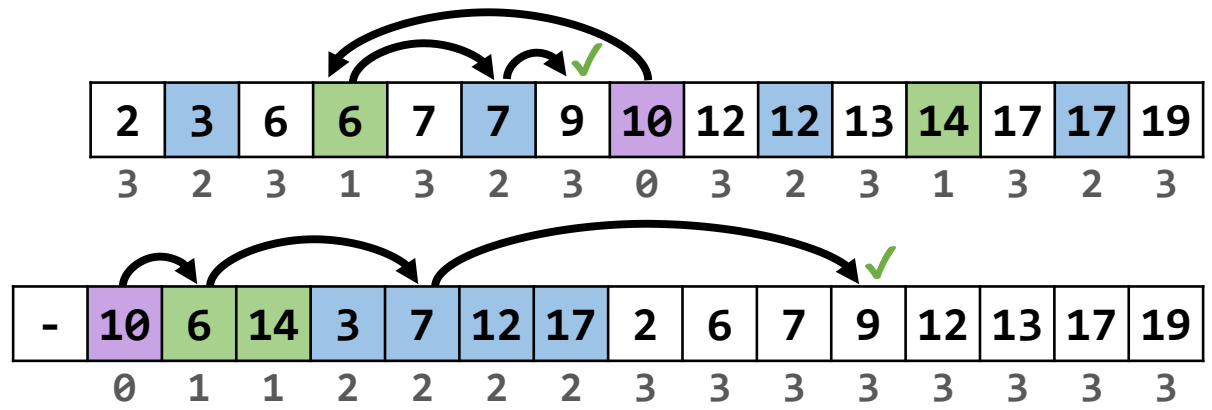

**Figure 6: The search path of traditional left-or-right binary search in comparison with Eytzinger binary search (EBS) for a point-lookup of key 9.**

On a GPU, we want to run multiple searches in parallel to fully saturate the compute capabilities: Each thread (identified by an integer `TID`) loads a single key to be searched, then performs one Eytzinger binary search in the reordered dataset, reads the associated row ID, and finally stores this result in a separate buffer.

## 4 Efficient Construction

We have discussed how to perform a lookup in Eytzinger order, but we still have to explain how to obtain this order efficiently. We mentioned before that there are several highly-optimized GPU sorting algorithms to construct an array in ascending order. However, the reordering that is required by **EBS** is traditionally specified as a single-threaded recursive algorithm [25]. Schlegel et al. [31] derive an explicit formula for the permutation required to reorder an ascending array into Eytzinger order, thus removing the recursion. However, they still assume the build process to be single-threaded, as their description hints towards keeping track of certain values while traversing the array.

In contrast, we believe that the inverse permutation, i.e., from Eytzinger order to ascending order, can be computed more cheaply on a GPU, in addition to not exhibiting any data dependencies between entries. This allows us to assign one thread to each slot in the yet-to-populate Eytzinger buffer. Each thread can then independently figure out which *source slot* in the ascending buffer to load from, and copy the contents into its assigned *target slot*.

We will provide our derivation for the inverse permutation in two steps, where we first consider arrays with sizes satisfying $n = 2^m - 1$, and then generalize the formula to arbitrary $n$ by adding a correction term.

If $n = 2^m - 1$, every level of the binary search tree contains the maximum number of nodes. We refer to such a tree as a *perfect* tree. Consequently, the first $l$ levels of this tree contain exactly $2^l - 1$ nodes. Therefore, a thread with `TID` $t$ can determine its tree level in the Eytzinger array by computing $l(t) = \lfloor \log_2(t + 1) \rfloor$. Here, level 0 refers to the topmost level. The position $i(t)$ within the level (starting at zero) can then be computed by subtracting the size of the previous levels, i.e., $i(t) = t - (2^{l(t)} - 1)$.

From Figure 6 we can see that the entries belonging to the bottommost level ($l(t) = m - 1$) in Eytzinger order are spaced apart with a stride of 2 in ascending order (colored in white), starting at offset 0. On the next higher level, the stride doubles to 4, and the entries in ascending order also start at offset 1 now (colored in blue). Continuing this pattern at every level yields the following formula: In thread $t$, the position of the element we have to load from the sorted array is given by

$$p(t) = \underbrace{2^{m-l(t)}}_{\text{stride}}\, i(t) + \underbrace{\frac{2^{m-l(t)}}{2} - 1}_{\text{initial offset}} = (2^{m-l(t)-1})(1 + 2\, i(t)) - 1.$$

To generalize this formula to arbitrary $n$, consider the array in Figure 7 with $n = 17$ entries. The tree representation of the search paths can never be a perfect binary tree, but by manually selecting the pivot elements for each step of the binary search, we can force the upper levels of the tree to be fully occupied, and the lowest level to be left-aligned. This is called a *complete* tree and is a prerequisite for linearization. We can compute the number of fully occupied tree levels using $m = \lfloor \log_2(n + 1) \rfloor = 4$, and the number $b$ of remaining lowest-level elements using $b = n - (2^m - 1)$.

If thread $t$ satisfies $t \geq 2^m - 1$, it is responsible for one of the $b$ lowest-level entries which end up in the last slots of the Eytzinger-reordered array, and those still have a stride of 2 in the ascending array (shown in orange). The remaining threads are assigned to upper-level entries, which are located between these lowest-level entries, until no such entries remain. Consequently, we have to skip the positions of the lowest-level entries when loading an upper-level entry. An entry that would be loaded from position $i$ if there were no lowest-level entries now has to be loaded from position $i + (i + 1)$, until $i > b$, in which case there are no more elements to skip. This fact can be added as a correction term, and we arrive at

$$p'(t) = \begin{cases} p(t) + \min(b, p(t) + 1), & \text{if } t \geq 2^m - 1, \\ 2 \cdot i(t), & \text{otherwise,} \end{cases}$$

using $t$ for the `TID`, $b$ for the bottommost level size, and $m$ for the number of full levels.

Note that all powers of two can be reduced to bit shifts, and all logarithms can be computed using a count-leading-zeros instruction, so the formula can be evaluated in constant time. In addition, two adjacent threads are extremely likely to process entries on the same tree level, and will therefore take the same branch during the computation of $p'(t)$, greatly reducing instruction divergence, which is a relevant factor to achieve good performance on a GPU.

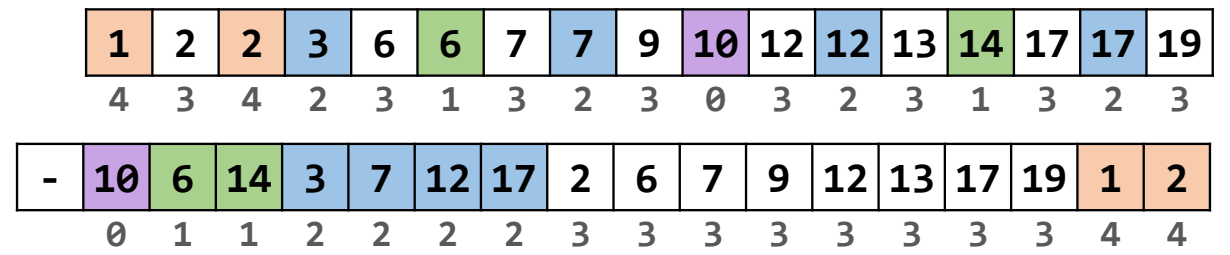

**Figure 7: Complete (but not perfect) Eytzinger array representation for $n = 17$ entries.**

## 5 Range Lookups

**EBS** has one important drawback: Range lookups and key duplicates are more difficult to handle since keys with neighboring ranks are no longer stored consecutively. Schlegel et al. solve this problem by deriving a formula to determine the in-order successor position within the linearized search tree. This allows them to iterate over all qualifying entries after locating the smallest qualifying entry.

However, their approach has two drawbacks: (1) For efficient iteration, the formula requires maintaining a base-$k$ representation of the rank of the currently visited entry in randomly-addressable memory, which not only suffers from a significant memory footprint when running up to 100000 lookups at once, but also introduces additional unstructured memory accesses, which is prohibitively expensive on a GPU. (2) When performing an in-order traversal, out of $k+1$ consecutively visited elements, $k$ elements are stored contiguously on the bottommost level, and the last element is an isolated entry in one of the inner nodes. In other words, for every efficient vectorized load on the bottommost level, we also have to perform one wasteful single-element access to a random location in the tree.

If we drop the requirement of having to visit the qualifying keys in order of their ranks, we can use a more efficient strategy that takes full advantage of coalesced loading on a GPU. The strategy is so broad that it can even be applied to **EBS**, which does not even utilize coalesced accesses by default.

Figure 8 shows an example for answering the range lookup [9, 12] on our usual key set in Eytzinger order (bottom part of the figure). While the qualifying entries no longer form a single coherent chunk, they are still consecutive within each Eytzinger level, so that a range lookup can be performed by a series of sequential scans, one for each level. The figure also shows that on each level, the start of this chunk of consecutive qualifying entries coincides with the search path we would take to look up the range query's lower bound.

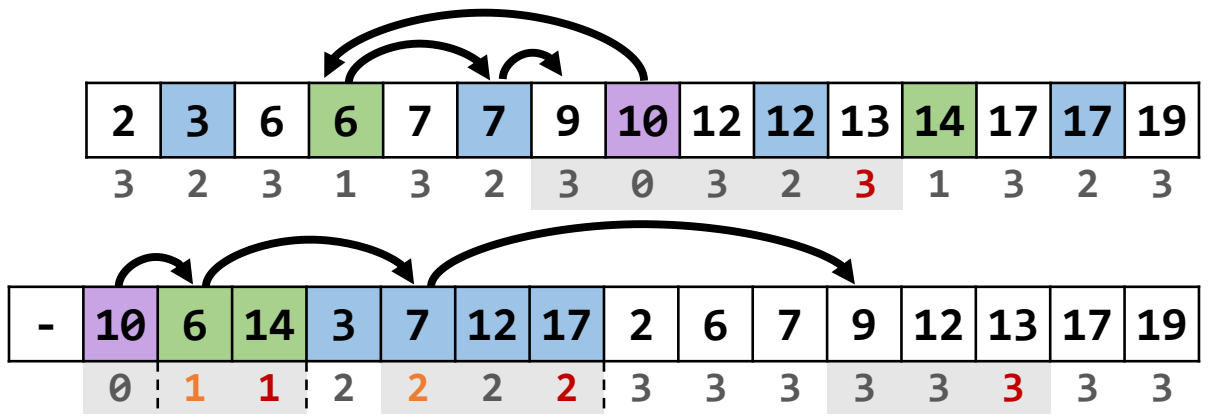

**Figure 8: Performing the range lookup [9,12] using Eytzinger order (bottom) and its equivalent in ascending order (top). Shaded level numbers mark the entries we have to load, numbers in red or orange denote entries outside the range.**

In the example, we find key 10 as the root entry, which is contained in [9, 12]. The search for the lower bound 9 now continues with the left child, which is 6. Even though 6 itself does not qualify for our query (indicated by the orange level number), the array entries directly to the right must be larger than 10 due to the search tree property, so we start our scan at the next position in the array. We continue until we reach the end of the level (shown by a dashed line), or until we find the first element exceeding the upper bound (level number in red), which is 14 in our case. Since 6 is strictly smaller than our lower bound 9, we can safely ignore the entire left subtree and continue with the right child, which is 7. We alternate between scanning a level and switching to the next level until we reach the leaf layer. Note that this implementation only visits up to two additional entries per level apart from the entries that are actually part of the range, which quickly becomes negligible as the size of the query range increases, resulting in minimal overhead.

### 5.1 Coalesced Scanning

As mentioned in the introduction, modern GPUs implement an optimization called *coalesced memory accesses*: If two neighboring threads read or write neighboring addresses at the same time, the hardware tries to replace the operations with a single, wider memory access. Coalesced memory accesses are the primary reason why scalable, high-performance GPU indexes are designed differently from CPU indexes.

The following micro-benchmark demonstrates the effect of coalescing: We allocate an array with $2^{28}$ 32-bit entries and fill it with random values. We then spawn a total of $2^{16}$ GPU threads, where $2^m$ neighboring threads (symbolically) form a group, resulting in $2^{16-m}$ disjoint groups. We then perform 1024 sequential lookup steps: In each step, each distinct group draws a random position $p$ within the array to load from, simulating diverging access paths for far-spaced accesses. Meanwhile, within a group, thread $i$ loads from position $p + i$, so that the group loads $2^m$ adjacent entries in total, and neighboring threads access neighboring entries. Note that the total amount of load operations remains constant across all choices of $m$, only their pattern changes. We plot the execution time of the lookup steps for various group sizes in Figure 9. The figure shows a 50% decrease in execution time each time we double the group size, up to 16. Moving to group size 32 yields a smaller benefit, and any increase beyond 32 does not result in faster execution. In other words, a coalesced load of 16 entries costs the same as loading a single entry, and loading 32 entries at once is cheaper than loading 16 entries twice.

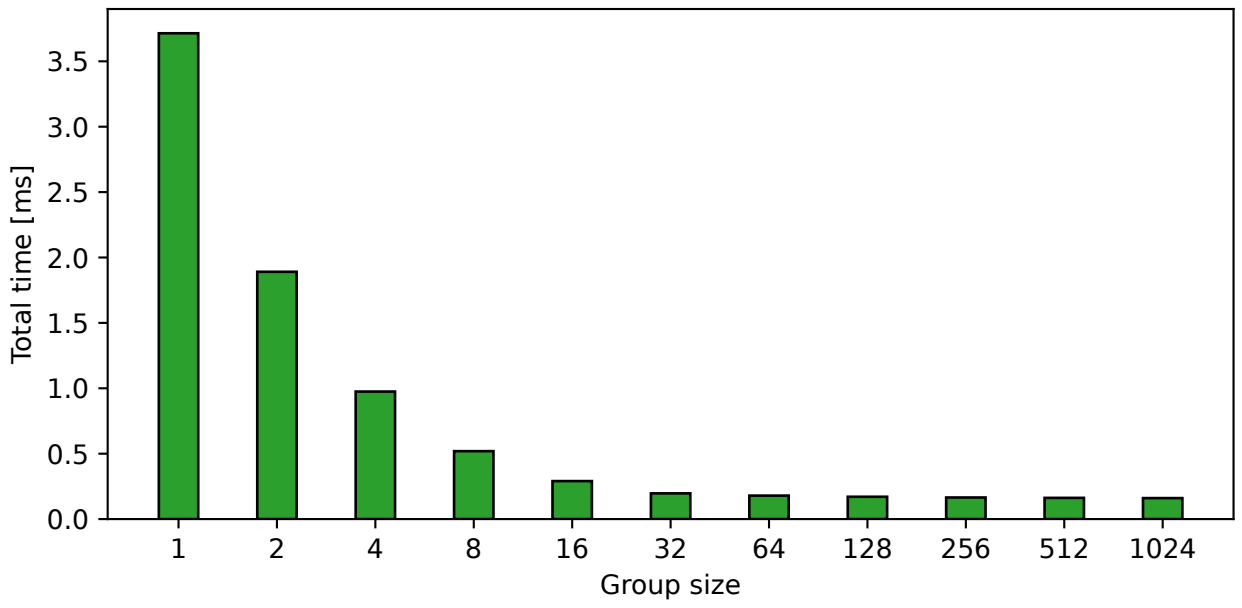

**Figure 9: Quantifying the benefit of coalesced memory loads.**

Since range queries exhibit a predictable memory access pattern of mostly loading neighboring entries, we can use coalesced loads to our advantage. Currently, two neighboring threads process two distinct range lookups, which likely target vastly different regions of the array, and therefore, the threads might load neighboring entries *sequentially*, but not *simultaneously*, so no coalesced loads take place. A better strategy is to form groups of $k \leq 32$ consecutive threads, which process a single range lookup collaboratively. This way, when scanning a level, we can load $k$ neighboring entries using a single coalesced access. Intuitively, this method of scanning is only beneficial if a sufficiently large number of entries has to be traversed in each level. Otherwise, the majority of threads in the group will load non-qualifying entries.

Conveniently, a specific property of the Eytzinger layout allows us to switch from one strategy to the other without having to switch back: If we find at least three qualifying entries within a single Eytzinger level, the amount of qualifying entries strictly increases on all subsequent levels, because the three entries have at least four qualifying child entries between them, and this argument can be applied inductively to the child entries. Therefore, if coalesced scanning improves performance on level $l$, it will definitely also improve performance on level $l + 1$.

We settled on the following hybrid strategy: In a group of 32 adjacent threads, each one processes a separate range lookup. Once a thread sees 32 qualifying entries on a single level, the thread stores its current scan state and pauses execution until all other threads in the group either complete their lookups or also pause. Afterwards, the threads communicate to determine which ones did not finish their scan. Then, sequentially for each unfinished scan, the scan state is scattered across all 32 threads using *warp shuffles*, and they continue the scan together with coalesced accesses. The threshold of 32 ensures that the switch from single-threaded to multi-threaded traversal never dominates the execution time. We will compare this strategy with the naive one in Section 8.4.

## 6 Eytzinger $k$-ary Search

Thread groups can also be utilized to speed up the search itself, but this requires us to generalize binary search to $k$-ary search ($k \geq 2$). This variant selects $k - 1$ evenly spaced pivots in each step, effectively partitioning the array into $k$ chunks. The target key is then compared to every pivot, and the search continues in one of the $k$ chunks depending on the outcome of the comparison. Naive $k$-ary search is uncommon on GPUs because it incurs even more random accesses than binary search (due to the pivots being spaced far apart), and its implementation is more complicated, all while the asymptotic complexity remains identical.

| 2 | 3 | 6 | 6 | 7 | 7 | 9 | 10 | 12 | 12 | 13 | 14 | 17 | 17 | 19 |
|---|---|---|---|---|---|---|---|---|---|---|---|---|---|---|
| 2 | 2 | 1 | 2 | 2 | 1 | 2 | 2 | 0 | 2 | 1 | 1 | 0 | 1 | 1 |

| 12 | 17 | 6 | 7 | 13 | 14 | 17 | 19 | 2 | 3 | 6 | 7 | 9 | 10 | 12 |
|---|---|---|---|---|---|---|---|---|---|---|---|---|---|---|
| 0 | 0 | 1 | 1 | 1 | 1 | 1 | 1 | 2 | 2 | 2 | 2 | 2 | 2 | 2 |

**Figure 10: Our previous dataset in ascending order (top) and 3-ary (ternary) Eytzinger order (bottom).**

However, by linearizing the levels of the search tree, we obtain *Eytzinger k-ary search* (**EKS**), shown in Figure 10. This generalization has two advantages over **EBS**: First, we can now load all pivots at once using a single coalesced load from a thread group (see Section 5) and then perform the comparison between the lookup target and the pivots in parallel. Secondly, since the thread group size can be chosen freely at runtime, it allows us to configure the trade-off between the height of the search tree and the node size, which is not possible on a CPU as SIMD instructions are mostly fixed-width. We will discuss this trade-off in the evaluation.

### 6.1 Build-phase Changes

We restrict the pivot count to a power of two for cache alignment reasons, which also allows us to drop the empty slot in the beginning of the array. Next, the formula for computing the source slots during build has to be generalized to arbitrary $k$, which can be done using similar arguments as before, ultimately arriving at

$$p'(t) = \begin{cases} p(t) + \min(b, (k-1)(p(t)+1)), & \text{if } t \geq k^m - 1, \\ i(t) + \left\lfloor \frac{i(t)}{k-1} \right\rfloor, & \text{otherwise,} \end{cases}$$

where the uncorrected position $p(t)$ is given by

$$p(t) = (k^{m-l(t)-1}) \left(1 + i(t) + \left\lfloor \frac{i(t)}{k-1} \right\rfloor\right) - 1$$

with

$$m = \lfloor \log_k(n) \rfloor, \qquad b = n - (k^m - 1),$$

$$l(t) = \lfloor \log_k(t) \rfloor, \qquad i(t) = t - (k^{l(t)} - 1).$$

We skip the details here for space reasons. Notice however that fixing $k = 2$ yields the equations proposed in Section 4.

## 6.2 Lookup-phase Changes

We propose two versions of Eytzinger $k$-ary search: The first variant **EKS (group)** performs each lookup using a group of $k-1$ threads. During each step, each thread loads one of the $k-1$ pivots and compares it with the target key. The threads then combine their results using warp balloting to obtain the number of pivots that are strictly smaller than the target, which determines the child node to visit next. If the current node starts at position $i$, its $k$ children start at positions $ik+(j+1)(k-1)$ for $0 \leq j < k$. For point lookups, the search terminates as soon as one thread locates the target key, or until the offset exceeds the array size. For range lookups, we scan a level collaboratively until we find a non-matching entry or the end of the level, after which we proceed to the next level.

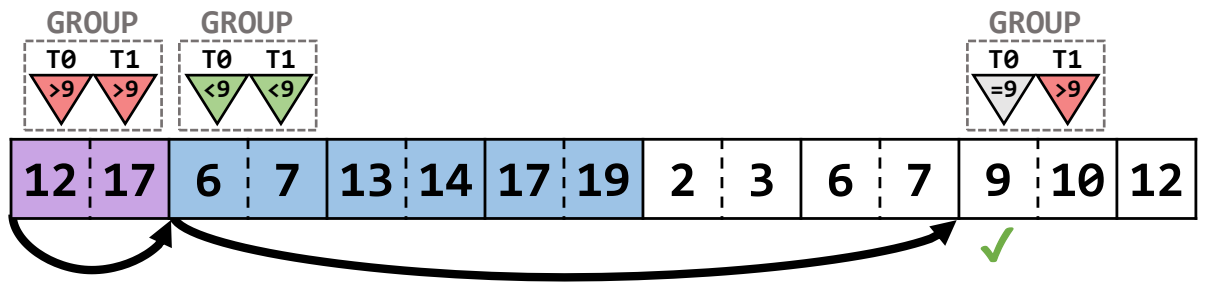

**Figure 11: Point lookup of key 9 using Eytzinger 3-ary search with a group of two threads.**

Figure 11 shows an example of how Eytzinger 3-ary search locates the target key 9 in three steps. Starting at the root node, two threads concurrently compare the two pivots 12 and 17 to the target key. Since no pivot is strictly smaller than 9, we proceed to child 0 at offset $0 \cdot 3 + (0+1)(3-1) = 2$. In this second-level node, both pivots (6 and 7) are strictly smaller than 9, so the next node to visit is located at offset $2 \cdot 3 + (2+1)(3-1) = 12$. Finally, thread 0 determines that its assigned pivot 9 is equal to the search target, after which the search terminates.

The second variant **EKS (single)** traverses the same underlying structure, but using only one single thread per lookup instead of a thread group. Within a node, the pivots are stored in ascending order, allowing us to perform a simple binary search to find the number of pivots strictly smaller than the lookup target.

# 7 Micro-Optimizations

After having discussed the core methods, let us see which micro-optimizations can be applied to potentially further enhance them.

## 7.1 Memory Layout

So far, we have not discussed how we materialize the row IDs that are associated with the reordered keys. There are two main options here: Store the row IDs in a separate buffer in the same order (structure-of-arrays or **SoA**) or interleave the row IDs with the keys in one single buffer (array-of-structures or **AoS**). Both variants share the same memory footprint, but the choice of layout usually impacts performance, as the amount of memory that must be loaded during lookup processing can differ.

## 7.2 Limiting Concurrency

As mentioned in the beginning, the GPU distributes work at the granularity of thread blocks, where by default, a scheduler *dynamically* assigns thread blocks to SMs. The dynamic aspect of this is that the scheduler can schedule multiple thread blocks to a single SM if it has sufficient resources available.

However, to reduce the load on main memory and the caches, it can be beneficial to purposely limit the amount of threads running at any given time. Hence, in the following, we propose a *static* scheduling strategy, which only spawns a fixed number of thread blocks, namely one per SM. Then, given `NBTHREAD` as the number of threads per block and `NBLOCK` as the number of blocks, the thread with number `TID` becomes responsible for processing the lookups at positions `TID + i * NBTHREAD * NBLOCK` (for $i = 0, 1, \ldots$). This effectively replaces the GPU's dynamic scheduler with a static assignment between lookup tasks and cores.

Apart from the scheduling variant, the number of threads per block `NBTHREAD` likely impacts performance as well, and therefore, we always test multiple block sizes ranging from 128 threads per block (the number of cores per SM) to 1024 threads per block (hardware limit, 8× more threads per block than cores per SM).

## 7.3 Cache Pinning

Because all variants of binary or $k$-ary search access the same entries during the first few steps, keeping them in the fastest cache at all times should yield an observable performance benefit.

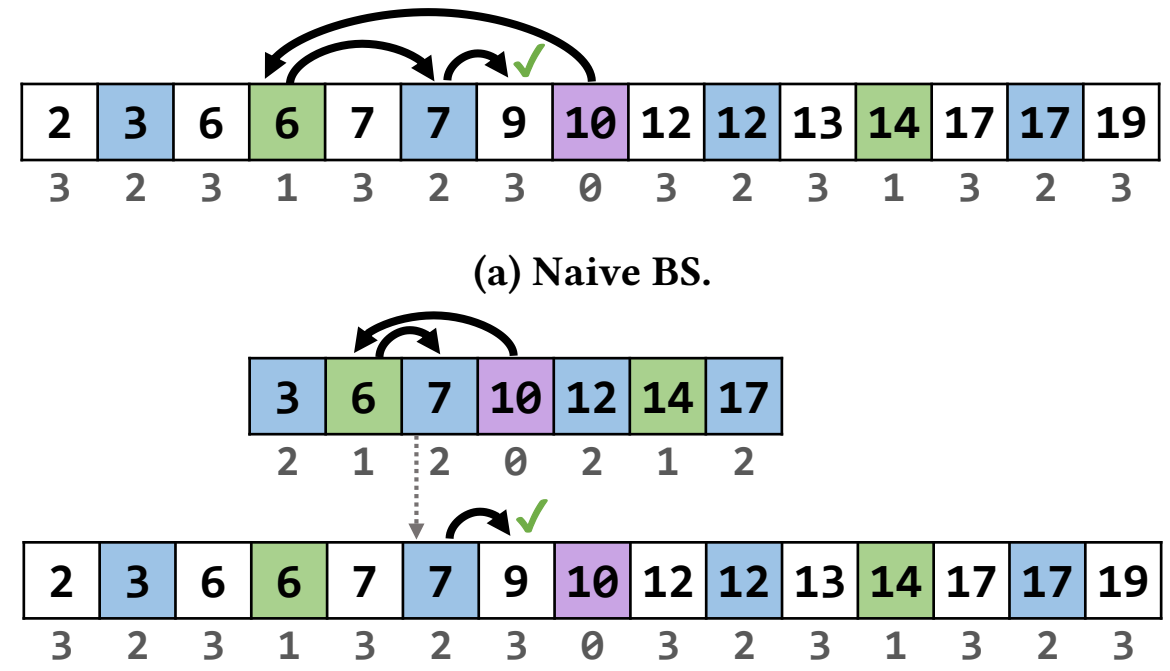

**(a) Naive BS.**

**(b) BS with cache pinning. The upper buffer is stored in scratch memory and hence pinned in the L1 cache of each SM.**

**Figure 12: Looking up key 6 with and without cache pinning. Orange arrows denote steps not taken.**

Unfortunately, by default, these entries may get evicted and replaced by entries down the search path due to the limited space in the cache. On a GPU, it is possible to actively pin entries by storing them in the up to 100 KB large scratch memory, which is carved out of the slightly larger SM-resident 128 KB L1 cache. By this, we can effectively exempt top-level entries from being repeatedly flushed from L1 cache during search. A similar approach appears in the implementation of [6], a GPU LSM tree implementation which will be one of our baselines.

As an example, consider Figure 12, which shows the **BS** point lookup of key 9 with and without cache pinning. Without cache pinning (Figure 12a), all four steps of the search are carried out in main memory. With cache pinning (Figure 12b) and a hypothetical 7-slot scratch memory region, we can carry out the first three steps required to find target key 9 entirely in scratch memory. Only after that, the search falls back to the equivalent slot in main memory, and continues with the last step.

Following the same motivation, we implemented cache pinning in **EBS**/**EKS** by just copying the leftmost slots of the Eytzinger array representation into scratch memory.

## 7.4 Local Lookup Reordering

Even with cache pinning, neighboring threads are still very likely to take different search paths after a few steps. However, if we minimize the absolute difference between lookup keys assigned to neighboring threads, their target positions will probably be close to each other in the sorted array, and therefore their search paths might diverge very late. An obvious way to achieve this is by globally sorting all lookup keys beforehand. Harmonia [33], a GPU B+-Tree, relies on lookup sorting as its primary source of speedup.

Unfortunately, globally sorting all lookups also has an unpleasant downside: Most GPU sorting algorithms operate out-of-place, which not only introduces additional memory capacity requirements, but also requires efficient allocation and freeing strategies in order to be feasible in practice. This fact is often ignored in the literature. A fast approximation of global sorting is *block-local sorting*, which sorts the key set directly from the threads' registers across all threads in a block. This variant, which we call *local lookup reordering* in the following, also runs much faster than sorting globally and only requires a small amount of SM-local scratch memory as a temporary buffer. Consequently, the sorting batch size is directly influenced by the number of threads in a block, up to the hardware limit of 1024 threads per block. Increasing the batch size without increasing the block size is still possible by pre-fetching multiple lookup keys into different registers of each thread, and then sorting across multiple registers. This approach is only limited by the number of available registers on the SM. In our experiments, we typically increase the register count as we decrease the block size to keep the sorting batch size constant.

Note that we expect the GPU to write the result of looking up the key at position $i$ into position $i$ of the result buffer. If we locally sort the keys, we still have to write the results in the original (unsorted) order, which now consists of almost exclusively random accesses. We can work around this issue by locally applying the inverse of the sort permutation to the results, thus restoring their original order before writing to the result buffer.

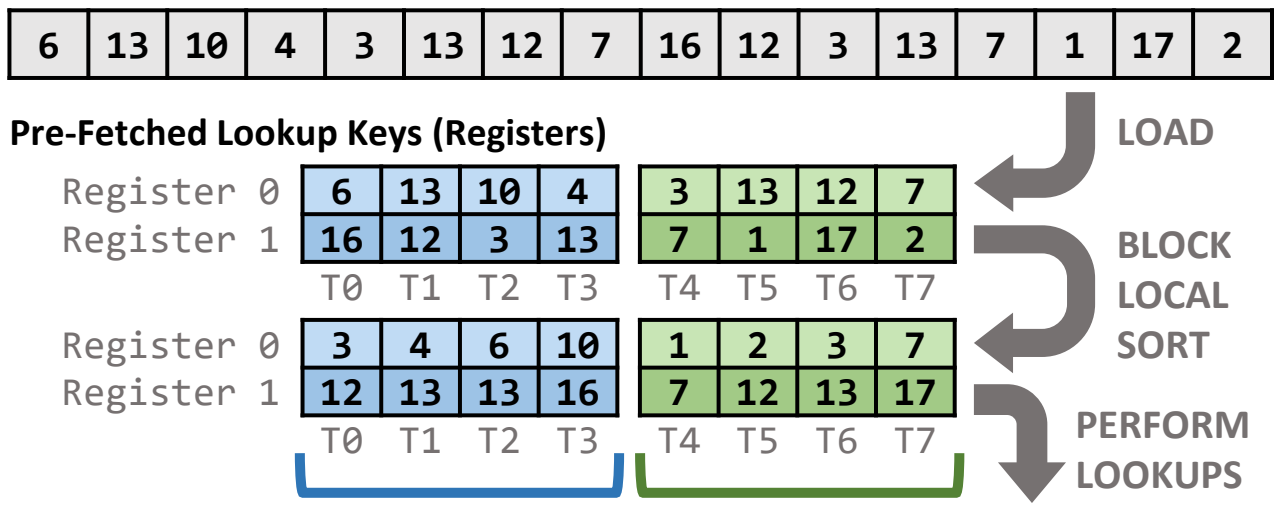

**Figure 13: Effect of reordering.**

To see how this sequence of pre-fetching and locally sorting works in detail, Figure 13 shows an example for 16 initially unsorted lookup keys. For simplicity, we assume that each block only has four threads, each of which allocates two registers for lookups. In the first step, we fill the two registers with the keys of the lookup array in the order they were stored. Then, we perform a block-local sort of the keys, resulting in a smaller absolute difference between neighboring keys. Afterwards, the lookup process proceeds as usual. The process for un-sorting the results works similarly.

## 8 Experimental Evaluation

In the following, we perform an extensive experimental evaluation which we split into two parts: First, we identify how to tune our proposed methods, also regarding the micro-optimizations. Second, we compare the best configurations against a set of state-of-the-art index structures, namely:

**B+**, a state-of-the-art GPU-resident B+-Tree by Awad et al. [8, 10, 17], without the overhead for multi-version support. Unfortunately, it only supports 32-bit keys and 32-bit values. Lookups are performed on nodes containing up to 15 keys with groups of 16 threads, performing a coalesced load for each node and determining the next child node concurrently during tree traversal. Leaf nodes are bulk-loaded to 100% of their maximum capacity to obtain the best possible performance.

**LSM** is a GPU LSM tree implementation by Ashkiani et al. [6]. Because the original code is several years old, we re-implemented the operations in our codebase. We fixed the base chunk size to $2^{16}$ bytes, which yielded the best performance across a range of sizes we tested. The search function uses a binary search with scratch memory caching for each tree level.

**PGM** denotes a PGM learned index [16], which is built using a sorted copy of the input dataset. For lookups, it receives a target key and returns an approximate position $p$ within the sorted dataset, with a configurable margin of error $\epsilon$, so that the target key's position is within $[p - \epsilon, p + \epsilon]$. A binary search can then be utilized to locate the exact position of the key. This is technically not a true competitor since no current **PGM** variant supports parallel construction on the GPU, but the lookup procedure can be moved to the GPU easily. We configured **PGM** as a single-level structure with $\epsilon = 64$.

**RX** refers to RTIndeX [18], which utilizes the hardware ray-tracing pipeline to implement point and range lookups. **RX** is an interesting baseline for this comparison because it visibly benefits from hardware acceleration for small and medium-sized datasets. Similar to our proposed structures, **RX** does not support any updates after the structure has been built.

**ST** is our own implementation of a static $k$-ary search tree with $k = 9$, stored in a single, contiguous array, inspired by CSS trees [30]. The bottom level contains all keys in ascending order. This is equivalent to **B+**, but does not require storing pointers to child nodes and replaces the leaf-level side links with a normal array traversal.

In terms of hashing schemes, we include all three mainstream variants of collision resolution: **HT (open)** refers to WarpCore, a state-of-the-art GPU hashtable [22, 23], which implements open addressing with cooperative probing to find target keys or identify empty slots. **HT (cuckoo)** is a non-resizable version of DyCuckoo [26]. This approach uses four arrays of blocks to store entries, and recursively evicts elements from overflowing blocks to resolve hash collisions. Lookups only have to check two blocks using thread groups, making them very efficient with regards to memory accesses. **HT (buckets)** is a slightly modified version of SlabHash [5], which uses a chain of linked nodes with 15 slots in each bucket. The number of buckets is determined by specifying the target node occupancy during build. Lookups also rely on thread groups to traverse the node chain of a bucket. Nodes are allocated by a rather sophisticated slab allocator, which had some hard-coded limits we had to remove in order to test larger build sizes.

In general, hashtable performance significantly depends on the load factor of the table, so we always test two configurations: One optimized for high performance, and another one optimized for a smaller memory footprint. We either report both or the better-performing one, depending on the experiment.

## 8.1 Experimental Setup

In all following experiments, unless specified otherwise, all data structures are built from a key set of $2^{28}$ unique and uniformly-drawn 32-bit unsigned integers (**B+** and some **HT** variants only support 32-bit keys). A subsequent lookup phase simultaneously probes $2^{25}$ (enough to saturate the GPU) uniformly-drawn keys from the build set in random order. This workload represents a batch lookup that would typically occur as part of a query pipeline that involves a sequence of index-based joins. Choosing a unique build set and drawing the lookup keys with a uniform distribution and in random order removes any chance of key ordering or dataset skew positively affecting cache efficiency and thereby influencing the measurements. Therefore, all results should be viewed as lower bounds on performance. We will conduct separate experiments to quantify the effect of skew and sortedness. For range lookups, we only run $2^{18}$ concurrent lookups that each scan a fixed-size range starting at a uniformly-drawn offset. The range size is chosen so that the query yields a pre-defined number of expected hits.

We assume all data and the index structures to be GPU-resident and therefore do not include memory allocations or CPU-GPU transfers in our measurements. These assumptions are common in both the database and the HPC community. We compile our implementation using version 12.8 of the CUDA toolchain.

## 8.2 Parameter Choice For Binary Search

We will start the evaluation by discussing the impact of limiting concurrency (Section 7.2), cache pinning (Section 7.3), and lookup reordering (Section 7.4), as well as good choices for the number of threads per block (Section 2) and lookup pre-fetching registers per thread.

It turns out that the impact of all parameter choices is greatly affected by whether the build set fits into the GPU's L2 cache. Therefore, we will present results separately for a build set size of $2^{15}$ keys (index size 256 KB) and $2^{28}$ keys (2.1 GB). Ultimately, we will propose a composite indexing strategy at the end of the paper to deal with this issue easily and elegantly.

Figure 14 shows a direct comparison between **BS** and **EBS** for a build set size of $2^{15}$ keys. We can make the following observations: (1) **BS** and **EBS** react similarly to parameter variation, and when using the same parameter configuration, **EBS** is always faster than **BS**. (2) Looking at the bars for 128 and 256 threads per block in

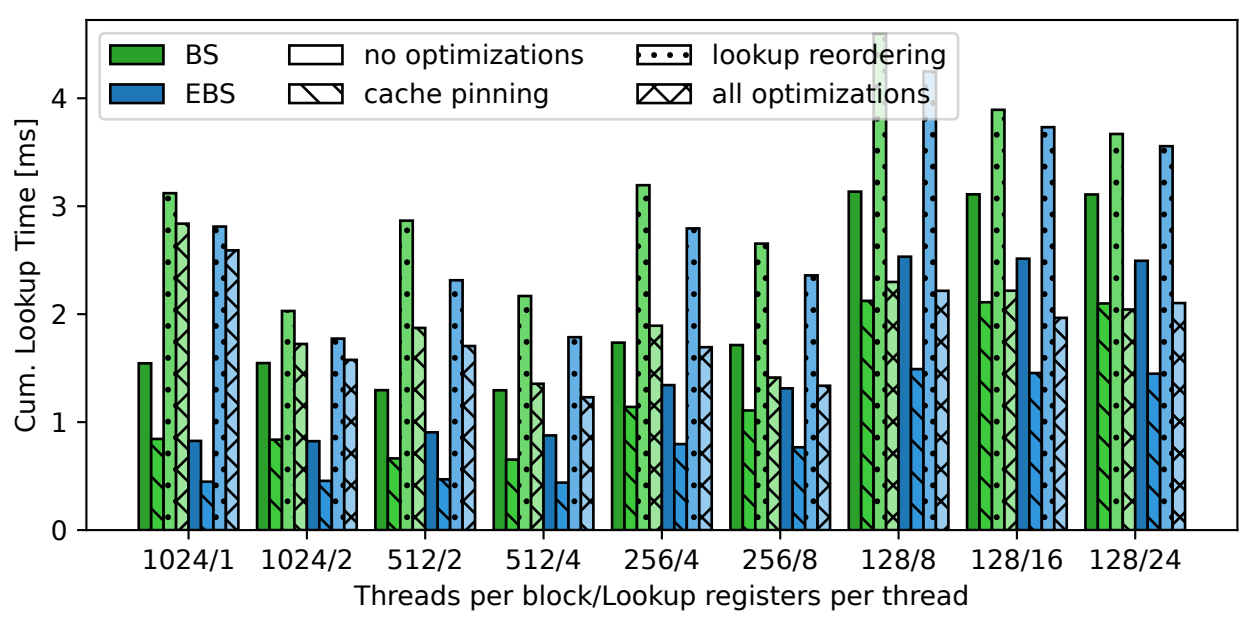

**(a) Limited concurrency.**

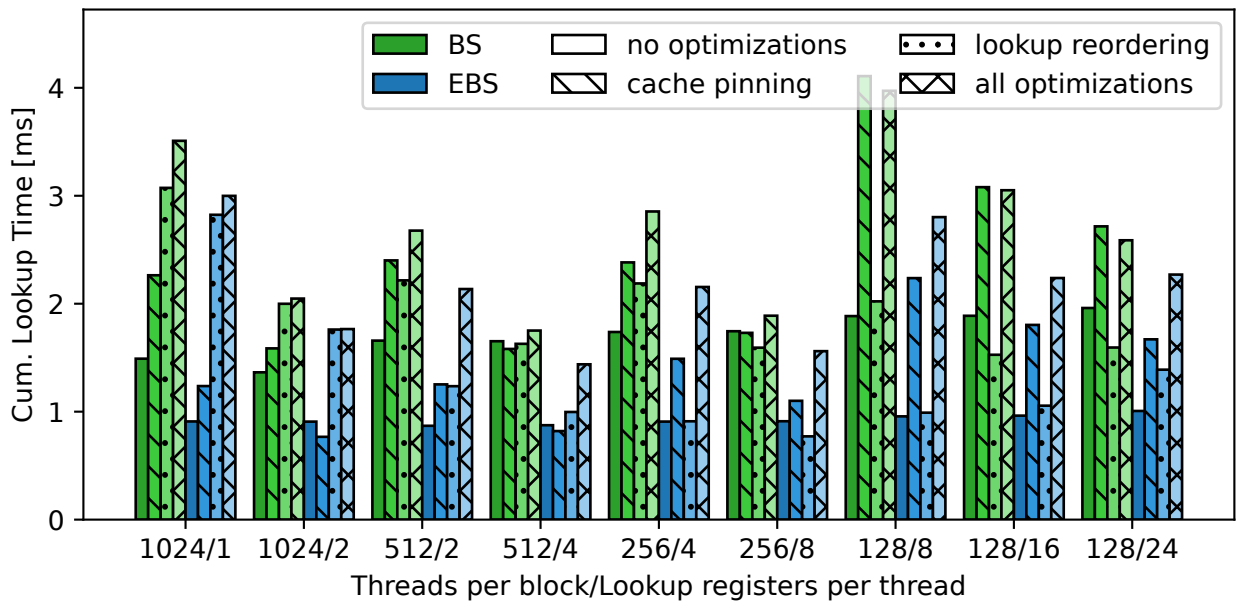

**(b) Unlimited concurrency.**

**Figure 14: Impact of micro-optimizations on BS and EBS for a build set size of $2^{15}$ keys.**

Figure 14a, we conclude that limiting the amount of concurrency too much significantly decreases performance. (3) However, limiting concurrency also opens up a sweet spot for cache pinning, which almost halves execution time when choosing 1024 or 512 threads per block. Even though the entire structure fits into L2 cache, pinning the upper levels to L1 cache yields additional performance benefits. (4) At the same time, cache pinning performs poorly when combined with the default GPU scheduler (Figure 14b). This is because the high amount of scratch memory required by pinning restricts the scheduler's ability to co-schedule multiple thread blocks to the same SM. At the same time, each newly scheduled block has to re-pin the upper levels from global memory, introducing additional overhead. (5) Lookup reordering mostly introduces unnecessary overhead at this small build set size. This is plausible, because the speedup for reordering originates from increased cache efficiency, which is already at 100%.

Figure 15 shows the same configuration for $2^{28}$ keys. The insights to be gained are more limited here: **EBS** still consistently outperforms **BS**, but the effectiveness of the optimizations differs. For example, cache pinning has little impact on **EBS**, but yields a significant improvement for **BS** most of the time. This can be explained by the fact that **EBS** places the first few search levels in a cache-efficient way by default, while **BS** has to rely on explicit pinning to achieve the same. In contrast, lookup reordering becomes very effective for both **BS** and **EBS** when combined with a low thread count and high register count. Interestingly, this significant improvement does not appear when also enabling the concurrency

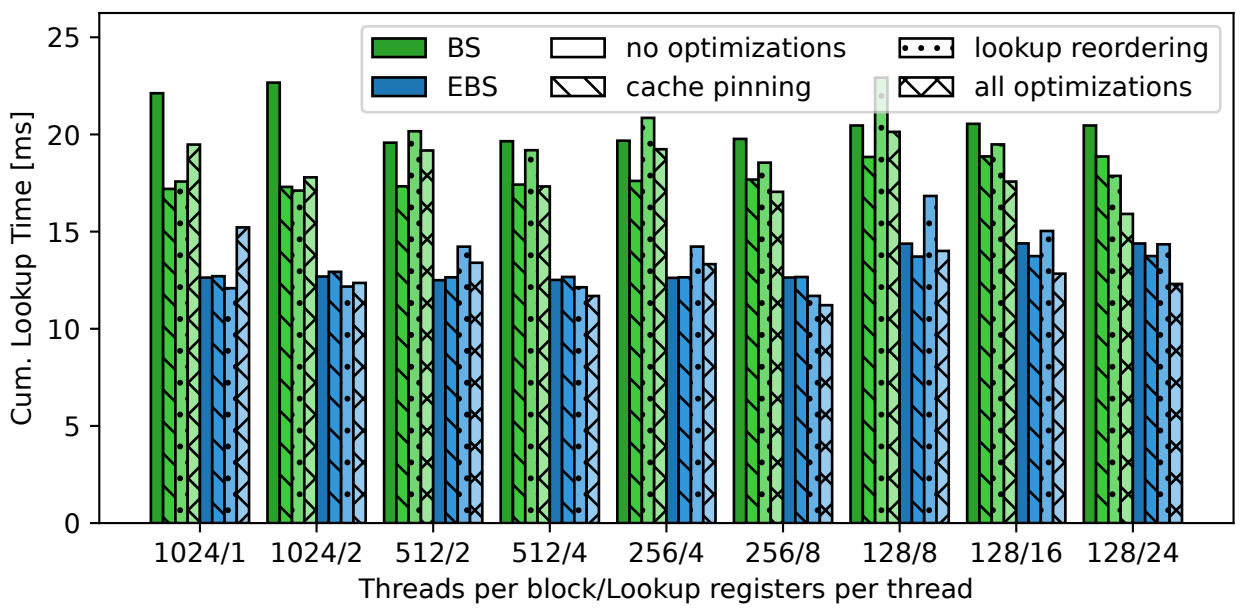

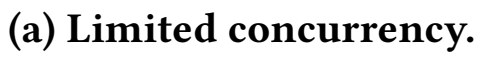

**(a) Limited concurrency.**

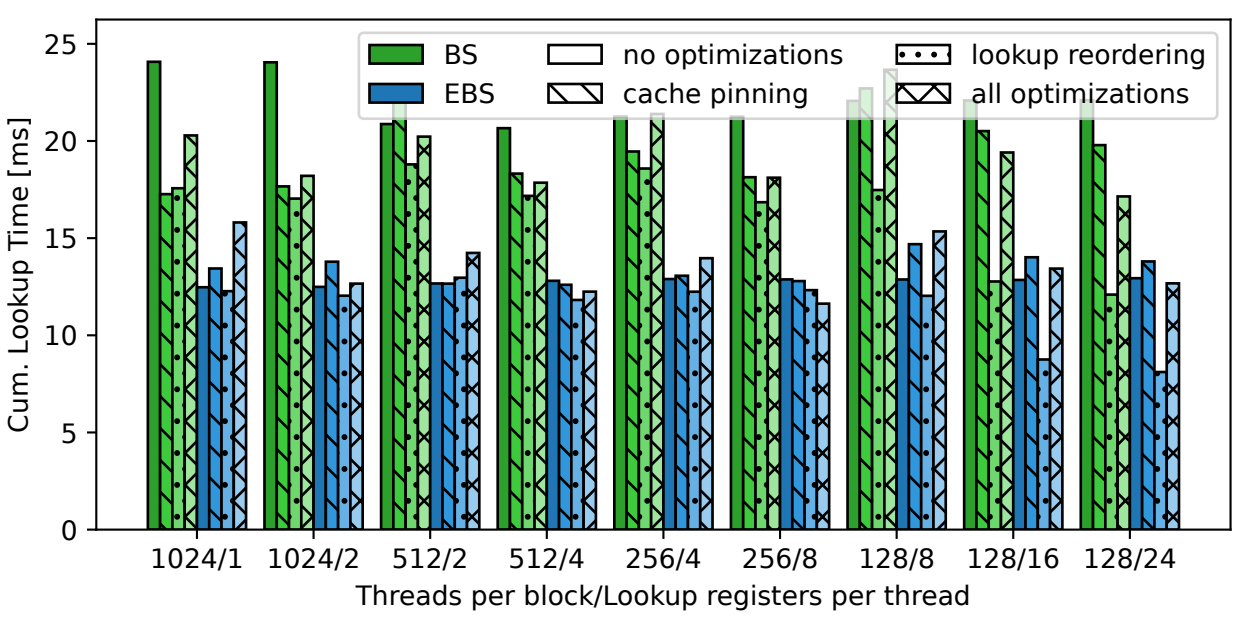

**(b) Unlimited concurrency.**

**Figure 15: Impact of micro-optimizations on BS and EBS for a build set size of $2^{28}$ keys.**

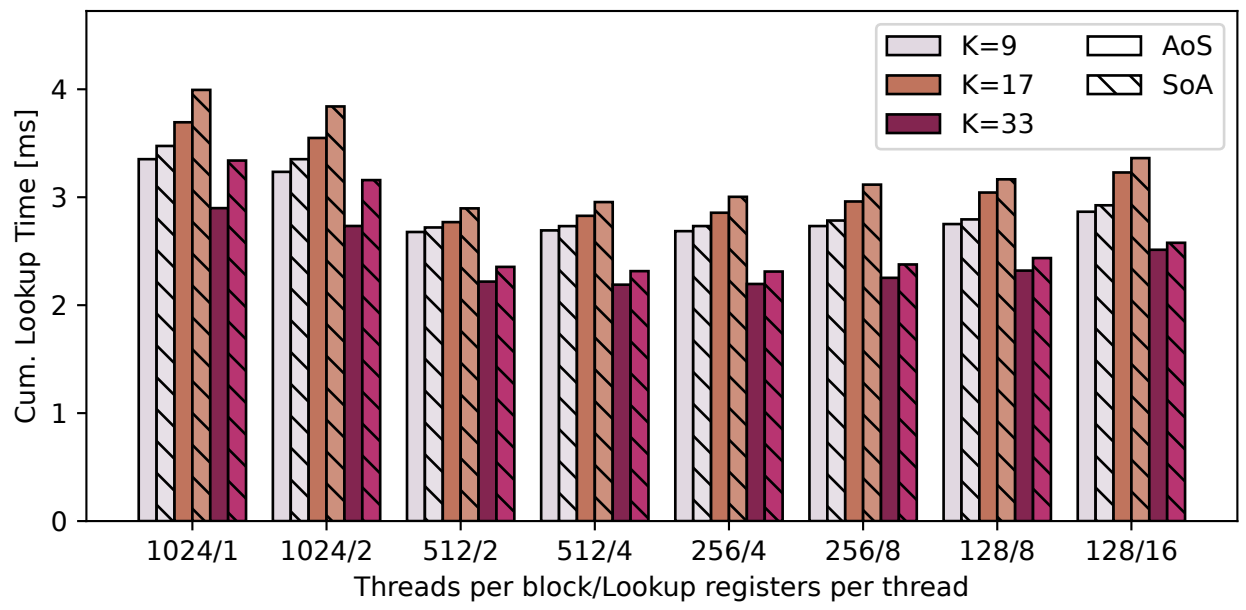

**(a) $2^{15}$ keys.**

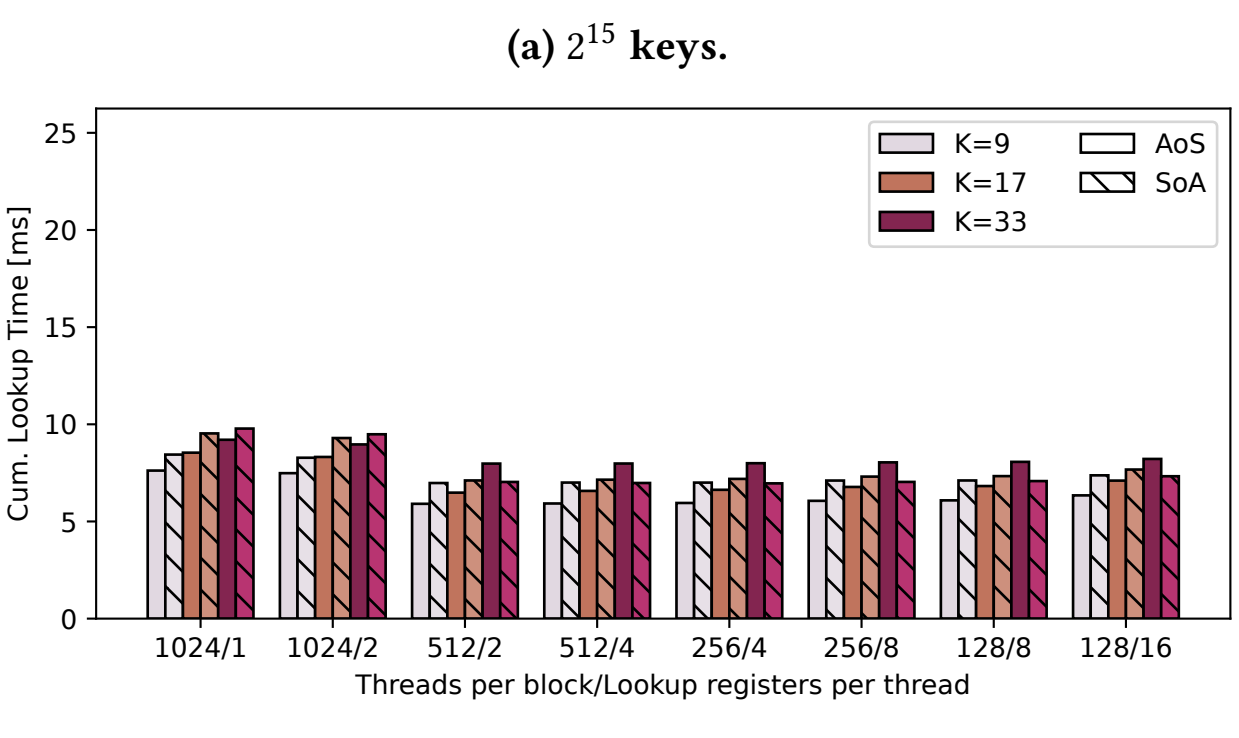

**(b) $2^{28}$ keys.**

**Figure 16: Impact of varying $k$ for EKS and for two build set sizes.**

limit, which appears to be the only major observable difference between Figures 15a and 15b.

Overall, we end up with two well-performing configurations of **EBS** that will be evaluated against a set of baseline indexes later on: (1) **EBS (small)**: 512 threads per block with 4 registers each, cache pinning, limited concurrency. (2) **EBS (large)**: 128 threads with 24 registers each, lookup reordering, no concurrency limits.

One thing we did not show so far is the **AoS**/**SoA** trade-off. This is because the choice of layout never made a significant difference in our point lookup tests. This is not the case for range lookups, as we will see later.

## 8.3 Parameter Choice For $k$-ary Search

Switching our focus to **EKS**, we first evaluate the thread group variant. Figure 16a shows the impact of varying $k$, the storage layout, and the thread/register count for a build set size of $2^{15}$ keys. Figure 16b shows the same for $2^{28}$ keys. We do not show the results of the concurrency-limited variants here because the performance was strictly worse compared the non-limited variants. The same applies to both cache pinning and lookup reordering.

Both figures show that choosing 1024 threads per block negatively impacts lookup times. Apart from that, cache-resident datasets seem to favor $k = 33$ (32 pivots per node) and **AoS** layout. However, when the dataset exceeds the size of the cache, this exact configuration performs worst among the alternatives. A smaller choice of $k$ might be counter-intuitive, but remember that $k$ also determines the size of the thread group used for traversal. Reducing $k$ increases the amount of thread groups working concurrently during lookups, which compensates for the additional search depth resulting from the lower fan-out.

Looking at the absolute numbers, we notice that **EKS** cannot be competitive with **EBS** on small inputs, but performs exceptionally well on large inputs. We therefore lock the configuration of **EKS (group)** to favor large build sets: **AoS** layout, $k = 9$, 512 threads per block with 2 registers each, and no concurrency restrictions.

All choices regarding the structure of the index, i.e., **AoS** layout and $k = 9$, will also apply when running **EKS (single)**. The remaining parameters are fixed where they maximize throughput under high-skew workloads, which we will introduce later in the evaluation. The concrete choices turn out to be unremarkable, we therefore refrain from showing any additional figures: No optimizations, no concurrency limits, 256 threads with 4 registers each.

## 8.4 Efficacy of Coalesced Scanning

Finally, we want to test the efficacy of coalesced scanning during **EBS** range lookups (Section 5). In Figure 17, we vary the number of expected hits per lookup between 4 and 14, and compare a strictly single-threaded scan against our proposed hybrid strategy that switches to thread groups after finding 32 qualifying entries in a single level. We also test both **AoS** layout and **SoA** layout. We divide the cumulative lookup time by the number of expected hits to avoid using a logarithmic scale.

The hybrid strategy clearly outperforms single-threaded scanning as the number of hits increases by a factor of up to 6.43, without incurring any measurable overhead for smaller ranges. Regarding layout, this experiment mirrors the previous result that the group-based **EKS** variant prefers **AoS** over **SoA**. This can be explained by noticing that a range lookup has to access the row IDs for *all* qualifying keys, not just a single one. With **AoS** layout, the row IDs are stored immediately next to the keys, and therefore already in L1 cache when they are accessed.

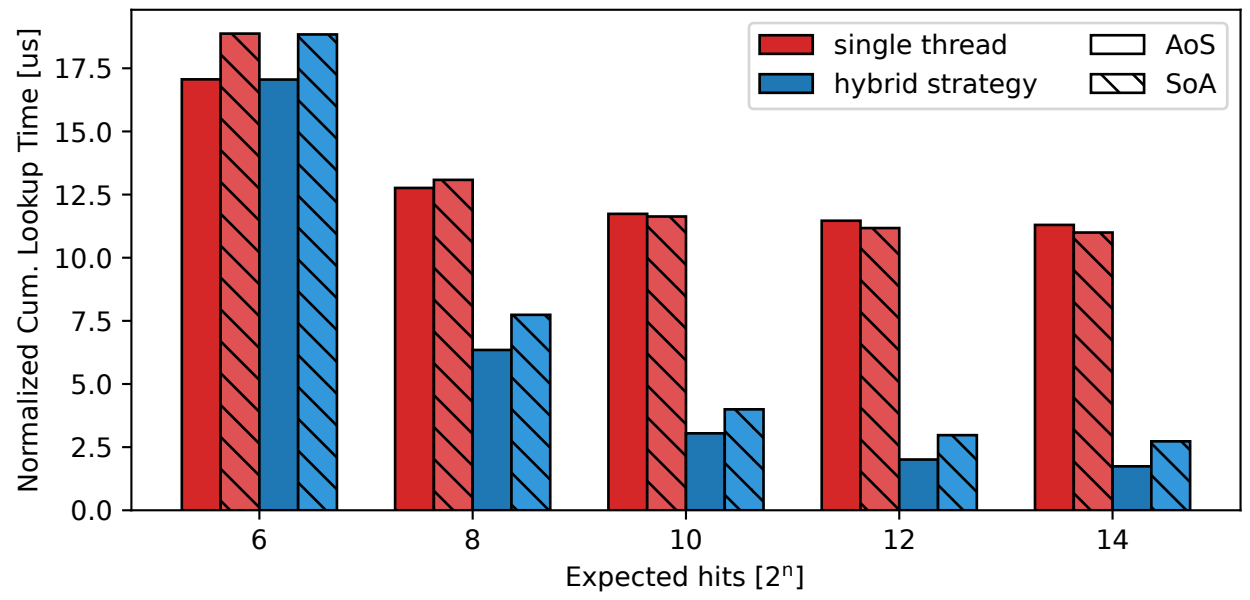

**Figure 17: Impact of coalesced scanning for EBS range lookups.**

## 8.5 Comparison with State-of-the-Art Indexes

We will now test whether **EBS** and **EKS** are competitive with the state-of-the-art GPU indexes. To that end, we compare the point lookup performance (Figure 18a), the time required for building each structure (Figure 18c), and the memory footprint (Figure 18b).

In addition to the naive text-book **BS**, we also show the best performance achievable though the optimizations discussed in Section 7 as **BS (opt)**. As such, **BS (opt)** serves as an upper bound for lookup performance. Similarly, we always show the best achievable **ST** performance as **ST (opt)**.

In terms of point lookup performance, the results yield several insights: Initially, **EBS (small)** outperforms all other indexes, even the **HT** variants. It is particularly noteworthy that **EBS (small)** performs lookups faster than **RX**, which defers the main data structure traversal to dedicated hardware (raytracing cores). At around $2^{24}$ keys, the index no longer completely fits into L2 cache. At this point, **EKS** achieves a similar or even better lookup time not only compared to **EBS (small)**, but compared to all other ordering-based indexes (i.e., everything except the hashtables). Meanwhile, **EBS (large)** remains competitive with **B+** and **BS (opt)**, but cannot match the performance of **EKS** and **ST** on large datasets.

In terms of build time, we can see that both **EBS** and **EKS** are competitive with **B+** and clearly outperform the **HT** variants. Building **EKS** is a bit faster than **EBS** because not only all writes are coalesced memory accesses, but also most reads. The small overhead over **BS** can be explained by the time required to rearrange the entries into Eytzinger order. Still, requiring well below 25ms to construct an index with $2^{28}$ entries shows that fully rebuilding our indexes in order to handle a batch of updates in read-mostly workload can be a viable option. Regarding memory consumption, as intended, the Eytzinger search variants show the smallest footprint of all methods.

## 8.6 Memory Efficiency

We also report a combined metric called *throughput per memory footprint*, which divides the throughput by the amount of memory the index permanently occupies, in Figure 19. Compared to raw throughput, this metric also rewards indexes for having a small memory footprint. Indexes that score high on this metric have a higher return on the memory invested.

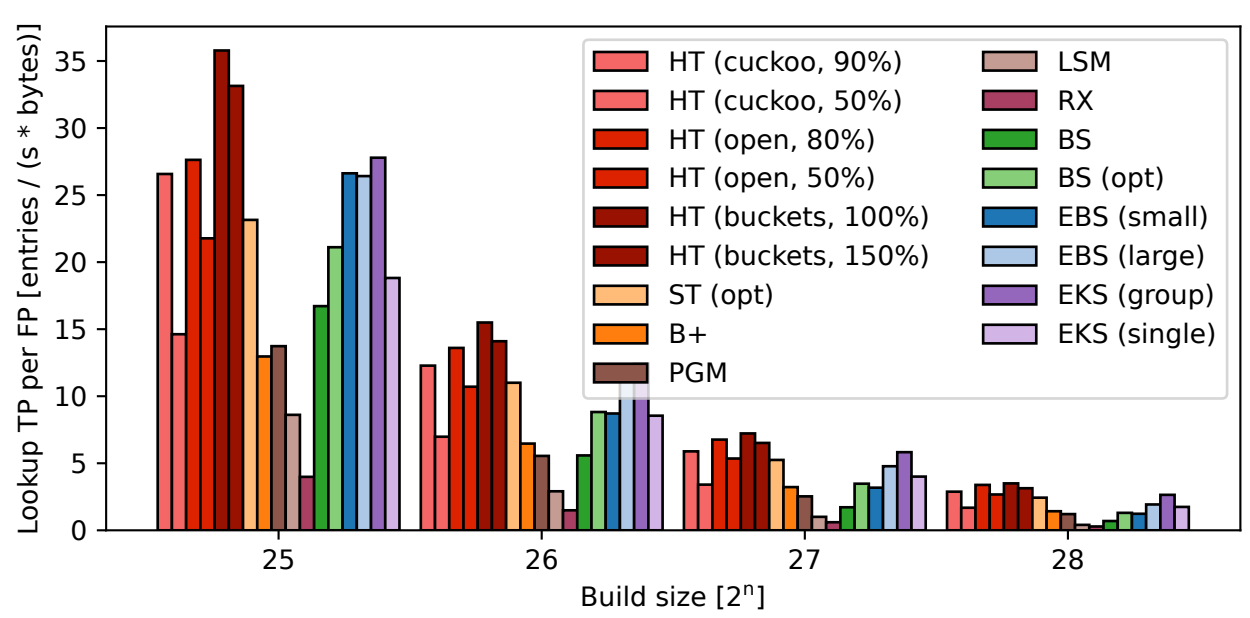

**Figure 19: Throughput per memory footprint.**

Even for large build sizes, **EBS** and especially **EKS** remain competitive with the hashtables on this metric due to their minimal memory footprints. Consistently, the performance gain of the hash tables over **EKS** is roughly proportional to the additional device memory they occupy.

## 8.7 64-bit Keys

In contrast to some of the baseline indexes, all Eytzinger search variants support 64-bit keys. We visualize the lookup performance for various sizes of 64-bit datasets in Figure 20.

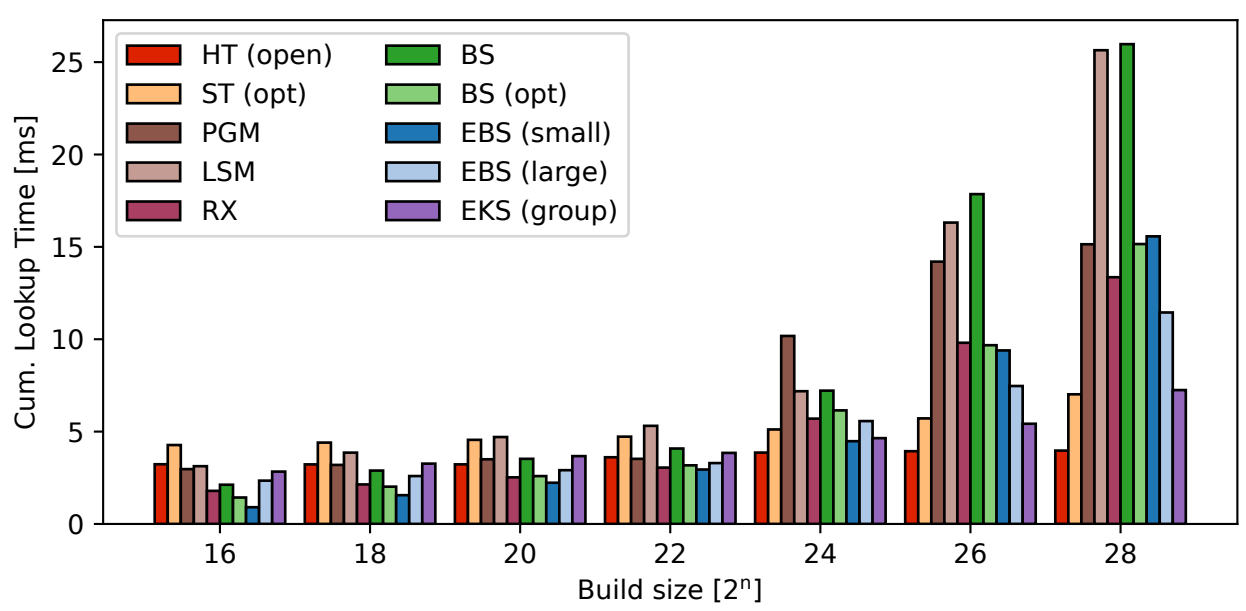

**Figure 20: Performance using 64-bit keys.**

The results remain consistent with the 32-bit experiment: **EKS** still outperforms all non-hashing baselines for large build sets, with **EBS (small)** yielding better executions times once the dataset fits into L2 cache.

## 8.8 Older-Gen Hardware

For the sake of completeness, we also measured the lookup time of all indexes on the older-gen NVIDIA RTX 4090 for various build sizes, with the results being shown in Figure 21. Interestingly, there now is a range of build sizes where **EBS (large)** performs best, but not by a large margin. Overall, the results remain comparable to

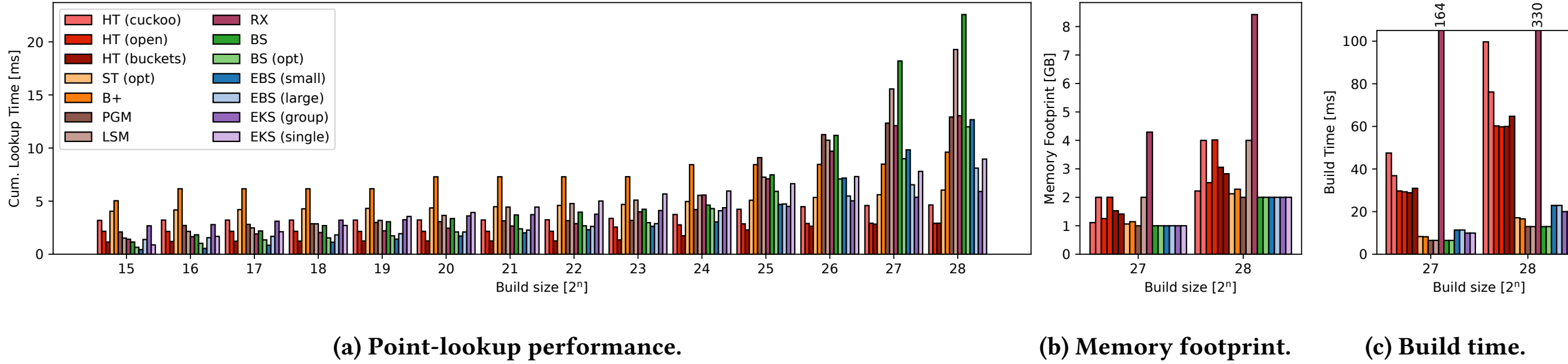

**(a) Point-lookup performance.** **(b) Memory footprint.** **(c) Build time.**

**Figure 18: Comparison of our optimized lightweight indexes with the baseline indexes while varying the build size.**

the RTX 5090, with **EKS** exhibiting even better lookup times on larger datasets.

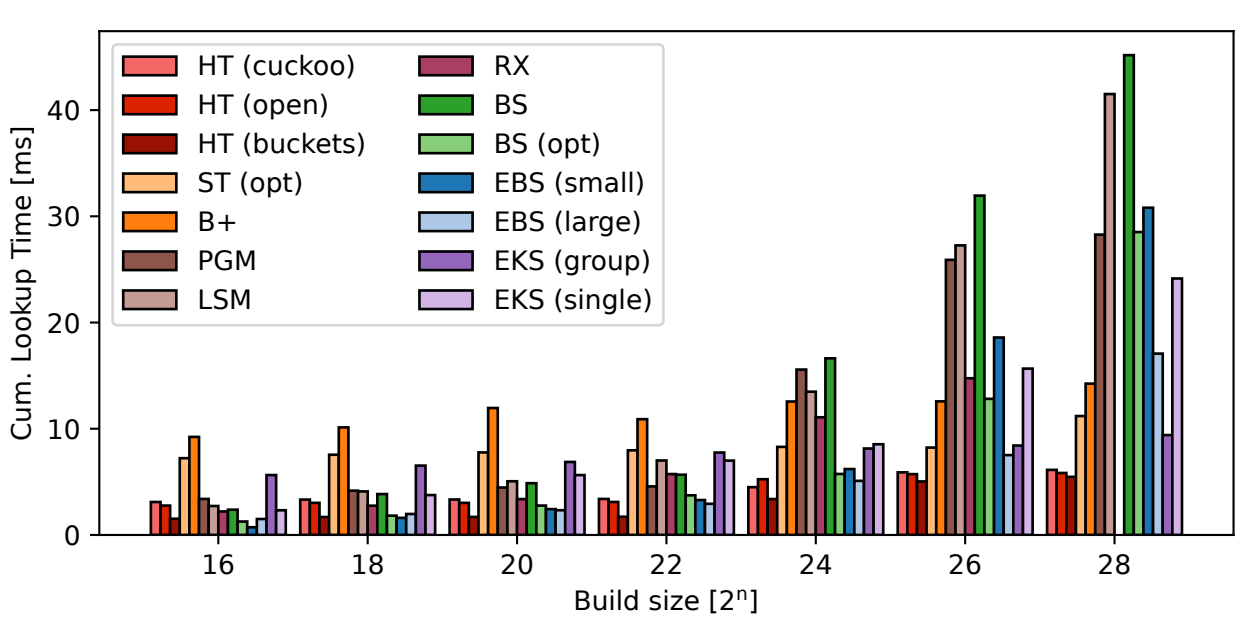

**Figure 21: Performance on RTX 4090.**

## 8.9 Lookup Skew

Next, let us consider a few edge cases to explore the limits of **EKS**: Figure 22 shows the lookup time of each index structure when the lookup keys follow a Zipfian distribution. We vary the Zipf exponent on the horizontal axis. The figure shows that **EKS** still outperforms all non-hashing index structures for small and moderate levels of skew. Once the exponent exceeds 1, the single-threaded lookup algorithms start performing significantly better, which also include **EKS (single)**. This effect can again be attributed to caching: At Zipf exponent 1.25, we only query 1201800 (0.44%) of the 268435456 inserted keys, amounting to 4.8 megabytes, which easily fits into the GPU's L2 cache along with the entries accessed during traversal. This mirrors the results from Figure 18a, where **EKS (group)** and **ST** work best when the dataset exceeds the size of the cache. At the extreme end (exponent 2.0), only 8016 (0.003%) of the build keys are being queried, resulting in **EKS (group)** performing poorly, but **EKS (single)** still handling this case gracefully. This shows that as long as we can estimate the level of skew in the lookup workload, no physical data reorganization is needed to retain optimal lookup performance with **EKS**.

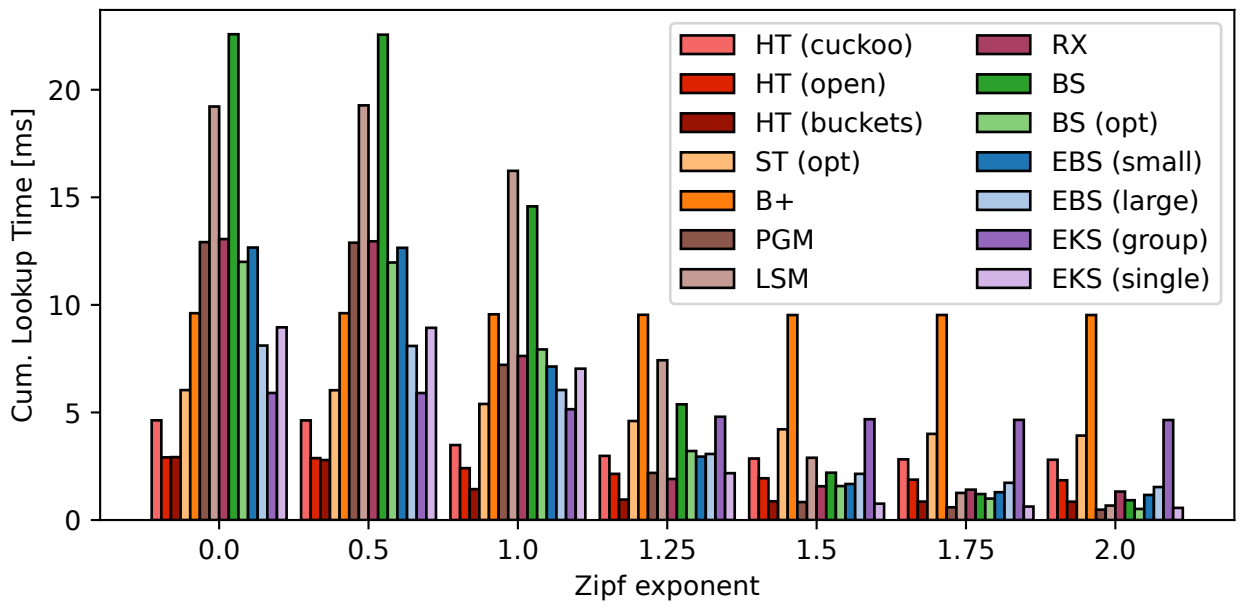

**Figure 22: Performance with skewed lookups.**

## 8.10 Pre-sorted Lookups

Another edge case occurs when the lookups happen to be ordered, shown in Figure 23. In this case, we query the vast majority of keys in the structure, but two adjacent threads are assigned similar target keys. This drastically improves lookup performance for single-threaded approaches because the 32 adjacent threads in a warp now take the same path through the data structure, which reduces the total amount of memory loads required.

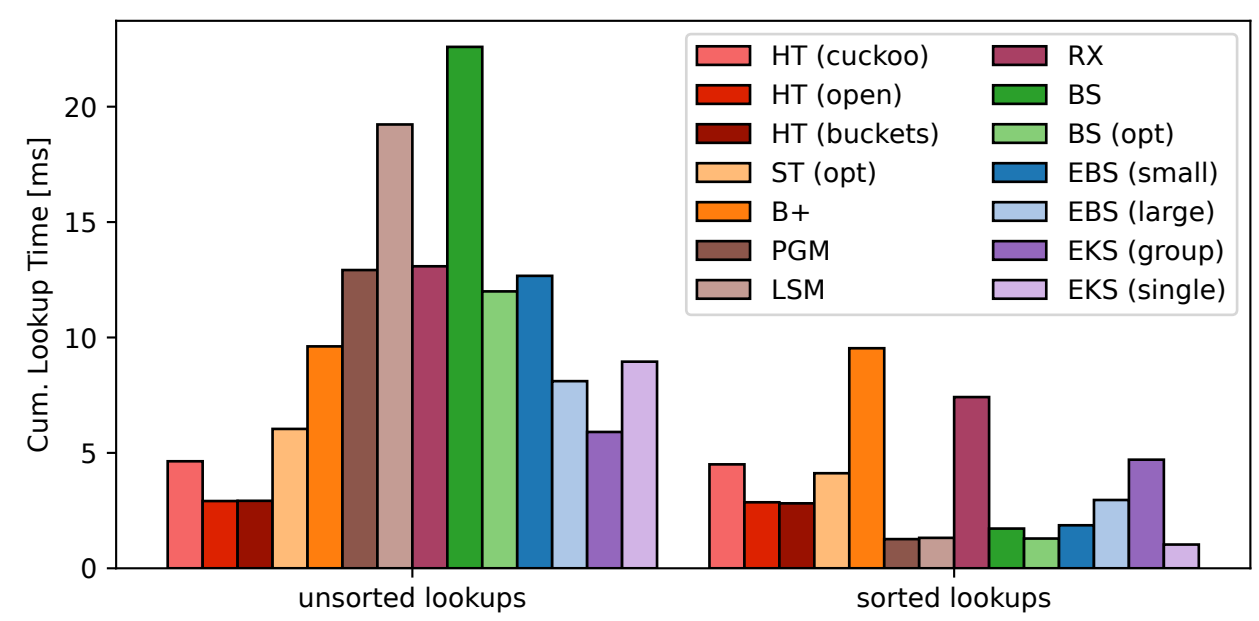

**Figure 23: Performance with pre-sorted lookups.**

Data structures relying on thread groups for traversal, i.e., **EKS (group)**, **ST**, and **B+**, conduct fewer simultaneous searches within a warp, and therefore benefit less from sorted lookups, making them the worst-performing approaches in this special case. Just like with high-skew workloads, **EKS (single)** again outperforms the best possible binary search here, allowing an easy fallback from **EKS (group)**.

### 8.11 Range lookups

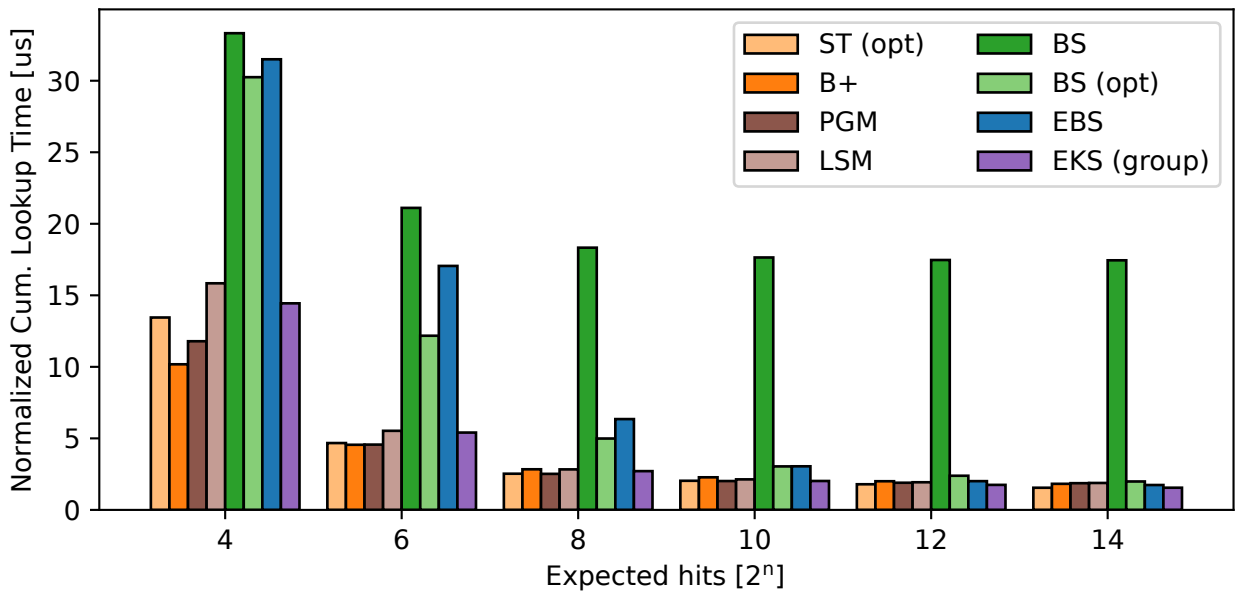

**Figure 24: Range-lookup performance.**

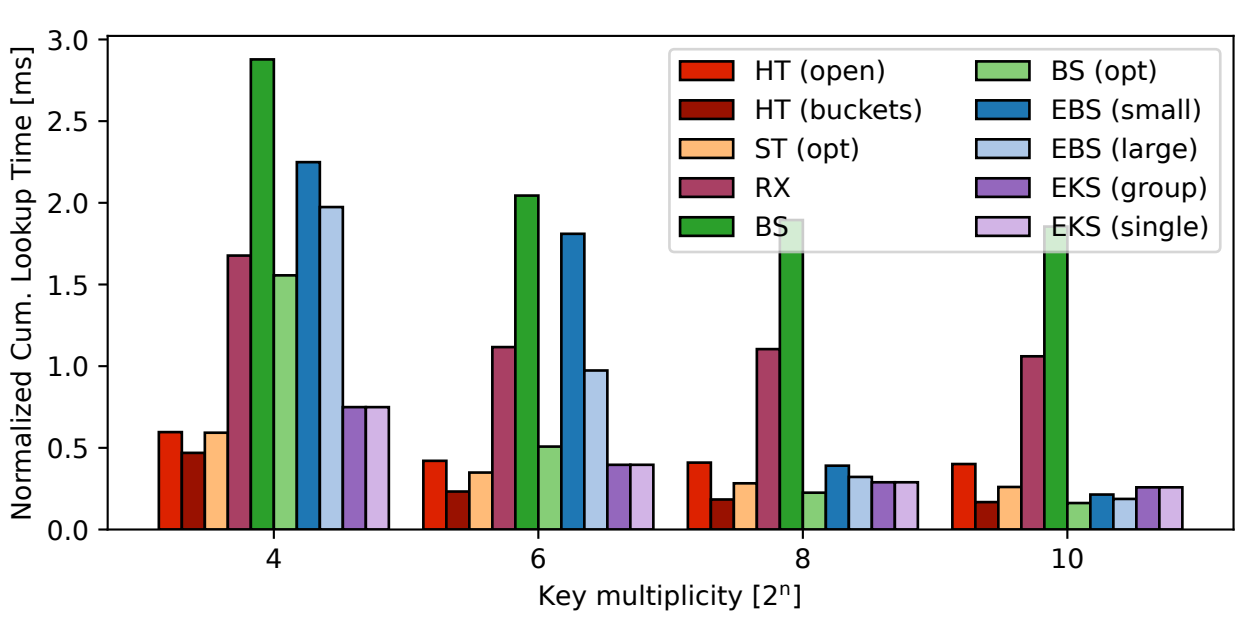

**Figure 25: Performance with duplicate keys.**

Finally, Figure 24 compares the range-lookup performance for various range sizes, again dividing the cumulative lookup time by the number of expected hits. **BS (opt)**, **PGM**, and **LSM** have been modified to also use coalesced scanning to serve as more competitive baselines. Even so, **EBS** and **EKS** outperform all competitors for sufficiently large ranges, while experiencing a minor slowdown for small ranges, which can be attributed to the unfavorable memory accesses that happen when only a few entries per Eytzinger level qualify for the query. We are confident that this issue can be solved with some additional tuning in the future.

Whenever the build dataset contains duplicates, all point queries have to be treated like range queries, as some amount of sideways traversal might be required to locate all occurrences of the target key. We investigated this special case separately in Figure 25, where each key in the dataset is replicated an identical number of times. Both **EKS** variants can handle replication just fine, but once we reach more than 256 replicas, the binary search variants perform slightly better. This can be explained by the fact that sequential scanning dominates the execution time at this point, which greatly benefits from larger thread groups. The node size for **EKS** determines the thread group size (8), while **EBS** relies on larger 32-thread groups. If this turned out to be a bottleneck in practice, one could either increase the **EKS** node size or switch to larger thread groups during traversal, similar to the hybrid range query strategy we proposed for **EBS**.

## 9 Related Work

In recent years, several CPU index structures have been ported successfully to GPUs. This includes hash tables [2, 3, 5, 20, 22, 26, 35], from which we selected our baselines **HT (open)**, **HT (buckets)**, and **HT (cuckoo)**, as well as comparison-based structures like our baseline **B+** [7, 9] and **LSM** [6]. Further, radix trees [1] have been proposed, where no public implementation is currently available how. There also exist GPU-resident spatial indexes such as R-Trees [29, 34] and GPU permutation indexes [28], which target different use cases.

Learned indexes typically tend to come with a high lookup performance in combination with a low memory footprint, but cannot be updated easily. Apart from the PGM learned index used in our evaluation, there is another learned index with GPU support [36], but unfortunately, its implementation is not publicly available. Previous work parallelized PGM lookups on a GPU[27], but the authors did not implement the subsequent binary search within the error bounds, which usually is the more expensive step.

If a static workload does not require efficient range lookups, perfect hashing places the keys in a collision-free hash table, which offers minimal lookup time at the cost of a more expensive build phase. One recent implementation [13] performs the build phase on the GPU itself, however, the lookup procedure requires prepartitioning on the CPU to work correctly. Because of this, we consider this index as a CPU index with optional GPU acceleration, and not as a competing GPU-resident index, which is why we did not include it in our evaluation. The pinnacle of perfect hashing is *minimal perfect hashing* (MPH), where the hash table is restricted to exactly as many slots as there are input keys, for optimal space efficiency. In contrast to **EBS** and **EKS**, which are also space-optimal, MPH does not store the keys explicitly, resulting in further memory savings. The build process is incredibly expensive, requiring 5 seconds to find an appropriate hash function for only 5 million keys on a GPU [11], and 90 minutes on a CPU.

## 10 Conclusion

In this work, we optimized several variants of binary search on a GPU and compared our implementation with sophisticated state-of-the-art GPU indexes, and conclude the following: If one is looking for an index that only supports point lookups, a hashing-based approach is still the most viable option. However, if range-lookup capabilities are required, one can use the following guidelines to obtain a peak-performance, space-minimal GPU index, which still matches the performance of hashtables on some workloads: (1) If the dataset to be indexed fits into the GPU's L2 cache, a parallel binary search on an array in Eytzinger order outperforms every other GPU index. (2a) If the dataset is larger than the L2 cache, traversing an array in $k$-ary Eytzinger order using groups of eight threads is the fastest indexing method. (2b) If the dataset is larger than the L2 cache, and the lookups are expected to exhibit high levels of skew or sortedness, $k$-ary Eytzinger order still yields the best results, but the traversal should be done using individual threads, not thread groups. Finally, larger range lookups should always utilize thread groups for traversing the set of qualifying entries, even when they are not stored contiguously. We proved that a run-time switch from single-threaded to grouped traversal is feasible if one cannot predict the number of qualifying entries.

Overall, we demonstrated that large and complex index structures do not necessarily lead to higher performance: The fastest static GPU index is also lightweight.